\documentclass[12pt]{article}
\usepackage{graphicx}

\begin{document}
\def\be{\begin{equation}}
\def\ee{\end{equation}}

\begin{center}
{\large\bf{Determination of the transverse width and distance of an object with a smartphone camera}}
\end{center}

\begin{center}
{Soumen Sarkar$^1$, Sanjoy Kumar Pal$^2$, Surajit Chakrabarti$^3$ \\
$^1$Karui P.C.High School, Hooghly, WB, India\\
$^2$Anandapur H.S. School, Anandapur,Paschim Medinipur,WB,India\\
$^3$Ramakrishna Mission Vidyamandira, Belur Math, Howrah, WB,India\\}
E-mail: tosoumen84@gmail.com; sanjoypal83@gmail.com; surnamchakrabarti@gmail.com
\end{center}

Credit Line: The following article has been accepted by The Physics Teacher. After it is published, it will be found at https://doi.org/10.1119/5.0065457.

A smartphone is a powerful learning aid in the hands of a large section of students around the world. The camera of the phone can be used for several learning purposes apart from its obvious purpose of photographing. If the focal length of the lens of the camera can be determined, several experiments in optics can be performed with it. In some  recent works,$^{1,2}$ the method for determination of the focal length has been discussed. When a real image of an object is formed by a lens of known focal length, one can determine either the distance or the transverse magnification of the object if the other is known. In this work we have shown that we can determine both the transverse size and the distance of an object, by photographing it from two  positions, separated by a distance along the line of sight of the camera. In a few other works,$^{3-8}$ the smartphone camera  has been used for conducting experiments in optics.

\noindent
{\bf\large{Theoretical Background for determining the focal length}}

When an image is formed by a convex lens, we have the relation$^{1,2}$
\be\frac {1}{v}-\frac {1}{u}=\frac {1}{f}.\ee Here
$u$ is the object distance and $v$ is the image distance measured from the lens which is assumed to be thin. $f$ is the focal length of the lens and is positive according to our sign convention. When a real image is formed by a convex lens, object and image positions are on the opposite sides of the lens and the image is inverted. Transverse magnification produced by the lens is given by \be m=\frac{v}{u}=\frac{I}{O}\ee where $I$ and $O$ denote the transverse sizes of the image and the object respectively and the magnification is negative. 
From the above two equations we have \be m=\frac{1}{1+\frac{u}{f}}.\ee Inverting this equation we get \be f=\frac{u}{\frac{1}{m}-1}.\ee  While calculating $f$ using Eq.(4), $u$ and $m$ are negative according to our sign convention. So, \be f=\frac{\left | u\right |}{\frac{1}{\left | m\right |}+1}.\ee 

\noindent
{\bf\large {Determination of unknown distance of an object and its lateral width}}

From Eq.(4) we get \be u=f_c(\frac{1}{m}-1).\ee
If $u_1$ and $u_2$ are the  distances of the  object from two positions of the camera with which we photograph the same object with transverse magnification $m_1$ and $m_2$ respectively, we get \be D=f_c(\frac{1}{m_2}-\frac{1}{m_1})\ee where $D=u_2-u_1$.This displacement should be along the line of sight of the camera lens. In terms of the object and image sizes we can write \be D=f_c(\frac{O}{I_2}-\frac{O}{I_1})\ee where $O$ is the size of the object and $I_1$ and $I_2$ are the image sizes formed on the camera sensor. Hence we get \be O=\frac{D}{f_c(\frac{1}{I_2}-\frac{1}{I_1})}.\ee In Fig. 1 we show the lens configuration. The lens in the displaced position has been shown with dash-dot marks. In our convention $u_1$ and $u_2$ are negative and $f_c$ is positive. If $I_1$, $I_2$ are  negative quantities, Eq.(9) ensures that the object size turns out positive for all situations. For simplicity  in calculation we can write \be O=\frac{\left |D\right |}{f_c\left |(\frac{1}{I_2}-\frac{1}{I_1}\right)|}.\ee We find the size of the object $O$ using Eq.(10) by determining $I_1$, $I_2$  and the distance $D$ since we already know the focal length of the lens. Using Eq.(6) we get the distances of the object from the two positions of the camera when we put $m$ negative. For calculation of the magnitude of the object distance we write Eq.(6) as\be \left | u\right |=f_c(\frac{1}{\left |m\right |}+1).\ee

\begin{center}
{\large\bf Experimental Data:}\\
\end{center}

\begin{table}[ht]
\centering
\caption{Data for the focal length $f_c$  of the camera lens  (1 pixel=1.4$\mu m$).\\Object size is the distance between two marks on the ruler.}
\begin{tabular}{ccccccc}
\hline
Obs.&$u$&Object size&pixel&Image size&$m=\frac{I}{O}$&$f_c$\\
no.&(cm)&$O$(mm)&&$I$(mm)&&(cm)\\
\hline
1&10.0&30.0&953&1.334&0.0445&0.426\\
2&10.0&40.0&1263&1.768&0.0442&0.423\\
3&20.0&30.0&458&0.6412&0.0214&0.419\\
4&50.0&50.0&307&0.4298&0.00860&0.426\\
5&100.0&50.0&151&0.2114&0.00423&0.421\\
6&100.0&90.0&271&0.3794&0.00422&0.420\\
7&200.0&30.0&45&0.0630&0.00210&0.419\\
8&200.0&60.0&91&0.1274&0.00212&0.423\\
9&200.0&776.0&1180&1.652&0.00213&0.425\\
\hline
\end{tabular}
\end{table}

\begin{table}[ht]
\centering
\caption{Pixel data for the image sizes on the camera sensor from two distances:\\Object size = 7cm
; 1 pixel=1.4$\mu m$}
\begin{tabular}{ccccccc}
\hline
obs&$u_1$&pixel&$I_1$&$u_2$&pixel &$I_2$\\
no.&cm& 1&mm&cm& 2&mm\\
\hline
1&10.0&2235&3.129&50.0&429&0.6006\\
2&10.0&2235&3.129&100.0&211&0.2954\\
3&10.0&2235&3.129&200.0&106&0.1484\\
4&20.0&1074&1.504&100.0&211&0.2954\\
5&20.0&1074&1.504&200.0&106&0.1484\\
6&50.0&429&0.6006&200.0&106&0.1484\\
7&100.0&211&0.2954&200.0&106&0.1484\\
\hline
\end{tabular}
\end{table}

\begin{table}[ht]
\centering
\caption{Data for displacement D, Object width obtained using Eq.(10) and Object distances obtained using Eq.(11): Object size = 7cm }
\begin{tabular}{cccccc}
\hline
obs &$u_1$&D&Object width& Estimated($u_1$)& Estimated($u_2$)\\
no.&cm&cm&cm&cm&cm\\
\hline
1&10.0&40.0&7.05&9.93&50.0\\
2&10.0&90.0&6.96&9.81&99.9\\
3&10.0&190.0&7.01&9.88&199.8\\
4&20.0&80.0&6.97&20.0&100.0\\
5&20.0&180.0&7.02&20.1&200.0\\
6&50.0&150.0&7.01&49.7&199.8\\
7&100.0&100.0&7.07&101.4&201.5\\
\hline
\end{tabular}
\end{table}

\noindent
{\bf\large {Experiment and Results}}

We take the photograph of a ruler kept at a known distance as shown in Fig.2. The ruler is stuck on the wall with a double sided gum tape. The distance of the ruler from the camera can be measured by a tape as shown in the figure. By analyzing the photograph with Microsoft Paint software(`Preview' is the software to be used with Apple computers), we find the difference of pixel numbers at the two ends of a selected part of the image using the cursors of the software$^1$. This difference is   what we have termed pixel in Tables I and II.The length of this selected part, read from the ruler, gives the real width of the object. The idea about pixels can be found in greater detail in the article by Jesus J.Barreiro et al.$^9$  The length of 1 pixel was found, from the website$^{10,11}$ relevant for our camera. We call it $R$ and for our phone  $R$ was 1.4$\mu$m. The image width on the camera sensor is R$\times$ pixel. The ratio of the width from pixel reading and the real width gives the magnification $m.$ The focal length $f_c$ of the camera lens is obtained using Eq.(5).

In Table I, the second column shows the distance of the camera from the ruler.The pixel count corresponding to the distance selected on the image plane is shown in the 4th column and the real width of the selected part, is shown in the third column. Here we give the absolute values of the image sizes, magnifications and the object distances. We determine the focal length of the lens for different combinations of object sizes and distances. We get an average focal length $f_c=4.22\pm 0.01$mm. The  website$^{12}$  gives the focal length of our lens as 4.2 mm. We have worked with an Apple iphone, Model No. iPhone 12 mini. 
 
 In Table II we show the image sizes $I_1$ and $I_2$ on the sensor obtained from photographs taken from two distances  of the camera  $u_1$ and $u_2$ from the ruler. The real width of the object selected by the cursors of the software was 7cm. In Table III we show different values of $D=u_2-u_1$. Each row of Table II corresponds to the same row of Table III. Using Eq.(10) we determine the  width of the object which is shown in column 4 of Table III. We get the average width of the object as $7.01\pm 0.01$cm, which matches very well the nominal width of 7cm read from the ruler. We have estimated the magnifications $m_1$ and $m_2$ for $I_1$ and $I_2$ taking the object width as shown in column 4. Using Eq.(11) we estimate $u_1$ and $u_2$ which are shown in columns 5 and 6. Nominal values of $u_1$ have been shown in the second column. Nominal distances $u_2$ can be obtained by adding $u_1$ and $D.$ Both of them  match the estimated $u_1$ and $u_2$ well.

We get image sizes at micron level accuracy. We can neglect errors associated with them.The error in estimation of the object width arises mainly from the displacement $D$ and a small error in the focal length of the camera. A small uncertainty in the position of the lens inside the smartphone gets nullified by the displacement $D$ of the phone. So, $D$ gives the actual displacement of the lens. Thus, the accuracy in the estimation of the object width and its distances from the camera depends mainly on the accuracy of $D$. We get errors in the object sizes and $u_2$ less than about $1\%$. For $u_1$, error is within $2\%$. 

\noindent
{\bf\large{Minimum displacement $D=u_2-u_1$ required for arbitrary object size and distance}}

From Eq.(2) we get \be\delta m=\frac{\delta I}{O}.\ee Taking the minimum detectable change in image size on sensor = R, the minimum detectable change in magnification $\delta m_{min}$ will be,
\be\delta m_{min} = \frac{R}{O}\ee which is a dimensionless number. From Eq.(3) we get \be \frac{dm}{du}=-\frac{\frac{1}{f}}{(1+\frac{u}{f})^2}.\ee For $\left |u\right |>>f$ \be \delta m=-\frac {f}{u^2}\delta u.\ee
So the minimum shift in position before the second shot is taken  should be \be \delta u=\frac{u^2}{f}\delta m_{min}.\ee Here, we have shown only the magnitude of $\delta u$ as our displacement $D$ can be both towards or away from the object. 

We can illustrate this with a datum from Table III. In the seventh row an object of width 7cm is at a distance of 100cm from the camera. We can calculate the minimum distance that one should move before taking the second shot. We get the minimum shift as 0.47cm. However, we shift by a distance of 100cm which is large compared to the minimum. This is to get a larger pixel difference between the two shots keeping the absolute width of the smaller image sufficiently large in terms of pixel. 

The necessary minimum shift can be larger if the object is at a far off distance.
In order to find the height and distance  of a 50m high building at a distance of 1km, one has to move a minimum of $\delta u=\frac {10^6 m^2}{4.22 mm}\frac{1.4}{50\times 10^6}=6.6m$ assuming a clear camera sight. Eq.(13) has been used for $\delta m_{min}.$

Even without knowing the actual distance and height of the object one can guess them. 
One can find the minimum distances to shift from Eq.(16) by putting some plausible combinations of numbers for the distance and the transverse dimension. Then one can decide to move by a distance larger than the maximum that one calculates.The only thing that matters is that the movement should be large enough so that the pixel difference between the two images and the absolute pixel counts on the sensor are sufficiently large. An image measured on the sensor usually has about 3-4 pixel difference in repeated measurements due to the measurement error. The pixel difference and the absolute pixel counts from two positions should be large compared to this measurement error. 

If one measures the displacement $D$ accurately,  one should get an accurate measure of the distance and height of the object. Students may find it interesting to determine the distance and height of far-off tall trees, hillocks and similar other objects just by taking photographs from two positions with their smartphone cameras. Following the method described in this paper one may find alternative ways to standard methods of measuring large  as well as small length scales accurately.The method could be used in other experiments in the classroom. One can think of measuring the refractive index of water by photographing the apparent image of an object immersed in it$^{13}$.

\begin{figure}[h!]
\centering
\includegraphics[width=16 cm]{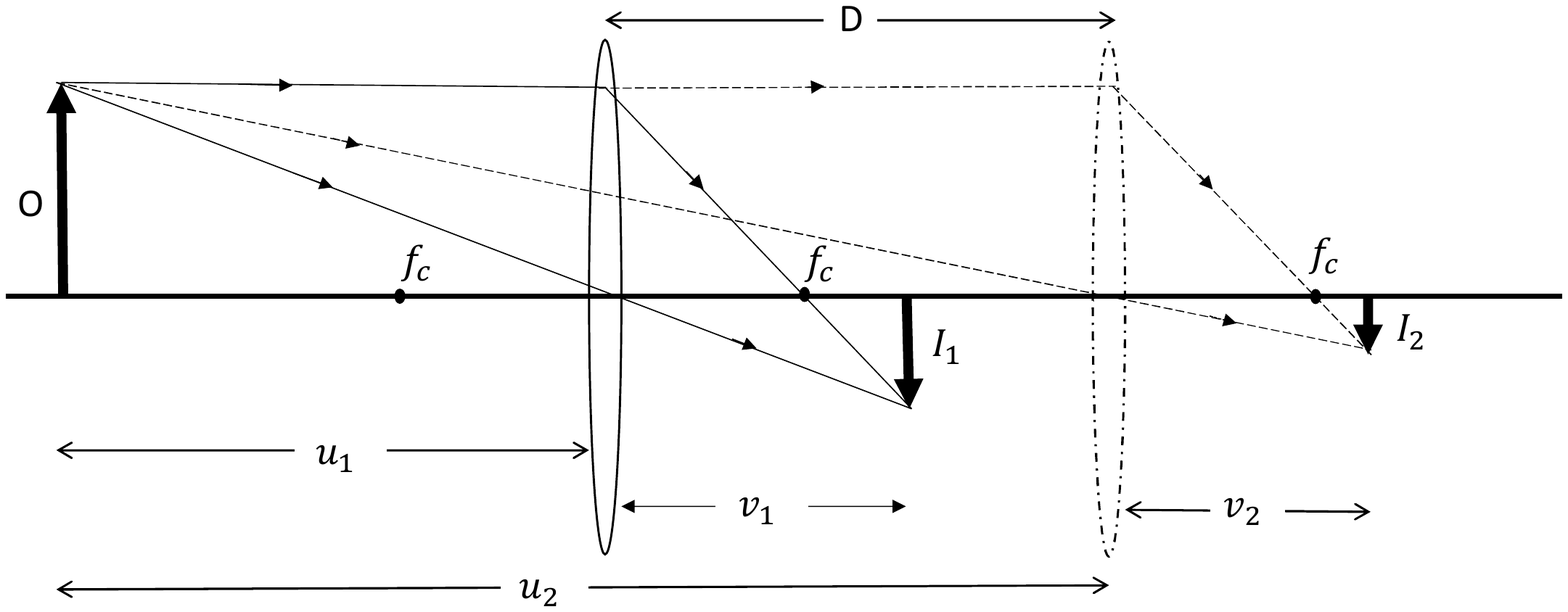}
Fig.1 {Image formation at the two positions of the lens}
\end{figure}

\begin{figure}[h!]
\centering
\includegraphics[width=16 cm]{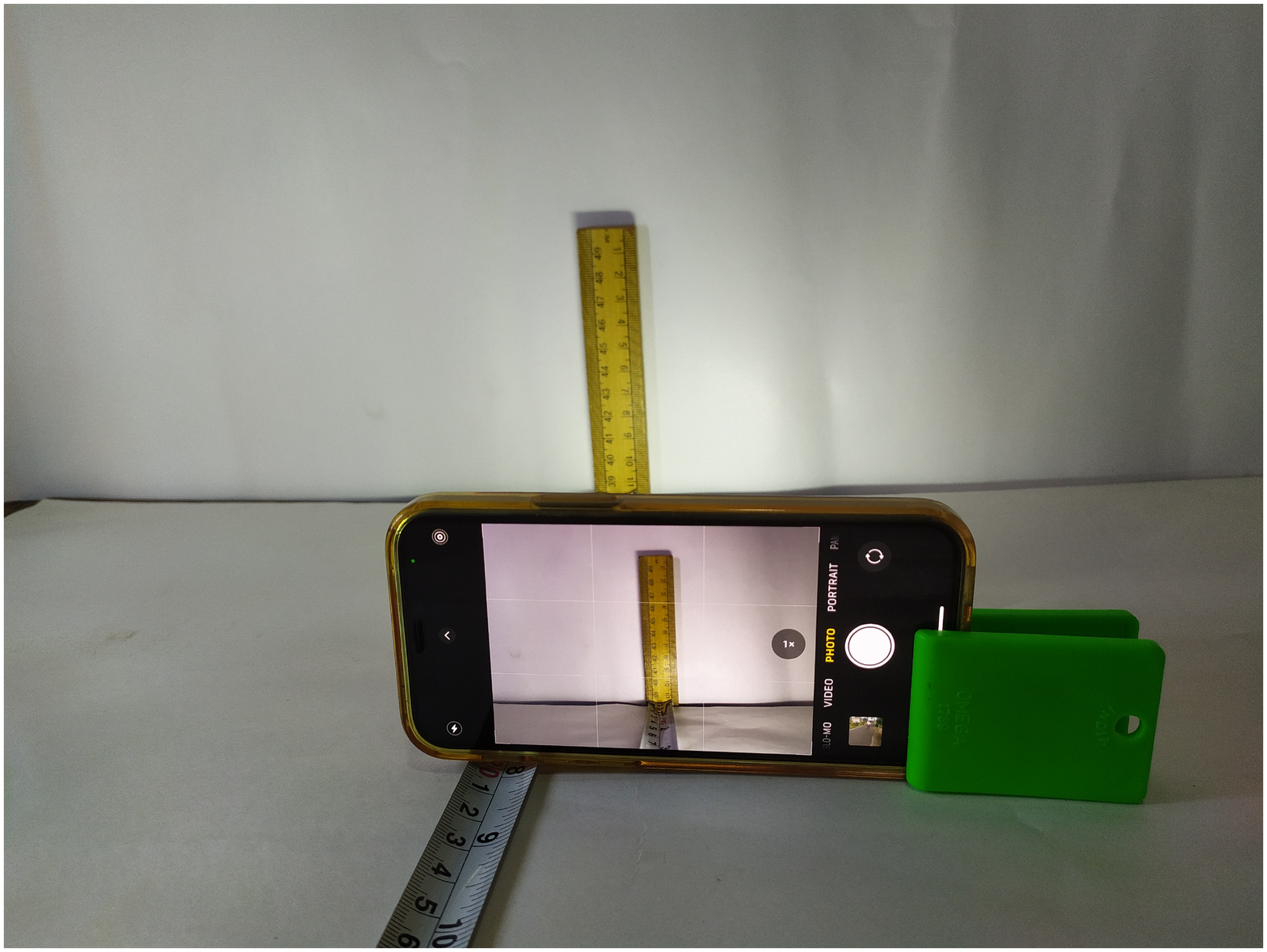}
Fig.2 {Image formation on the smartphone}
\end{figure}

\newpage
\begin{center}
References
\end{center}
\begin{enumerate}
\item Jun Wang and Wenqing Sun, ``Measuring the focal length of a camera lens in a smart-phone with a ruler,"Phys.Teach. \textbf{57},54 (2019).
\item Antoine Girot, Nicolas-Alexandre Goy, Alexandre Vilquin, and Ulysse Delabre, ``Studying ray optics with a smartphone." Phys.Teach. \textbf{58},133(2020).
\item Jack Freeland, Venkata Rao Krishnamurthi, and Yong Wang, `` Learning the lens equation using water and smartphone/tablets," Phys. Teach. \textbf{58}, 360 (2020).
\item Mickey D Kutzner, and Samantha Snelling, `` Measuring magnification of virtual images using digital cameras,"  Phys. Teach. \textbf{54}, 503 (2016).
\item Timo Hergemoller, and Daniel Laumann,``Smartphone magnification attachment:Microscope or magnifying glass," Phys.Teach. \textbf{55},361 (2017).
\item A.Pons et al.,``Learning optics using a smartphone," in ETOP 2013 Proceedings, (Optical Society of America,2013),paper EWP13.
\item Shangwen Chen et al.,``Single-image distance measurement by a smart mobile device," In IEEE Transactions on cybernatics, December 2017.
\item Suraphol Laotrakunchai et al.,``Measurement of size and distance of objects using
mobile devices," International conference on signal-image technology and internet-based systems (2013).
\item Jesus J.Barreiro et al., ``Diffraction by electronic components of everyday use," Am.J.Phys. \textbf {82} 257 (2014).
\item https://www.kimovil.com/en/apple-iphone-12-mini/camera 
\item https://www.gsmarena.com/apple-iphone-12-mini-10510.php
\item https://www.metadata2go.com/
\item Sanjoy Kumar Pal,Soumen Sarkar, and Surajit Chakrabarti,``Determination of the refractive index of water and glass using smartphone cameras by estimating the apparent depth of an object," arXiv:2111.06735v1 [physics.ed-ph] 11 Nov 2021
\end{enumerate}

\end{document}